\documentclass[preprint,amsmath,nofootinbib,twoside]{revtex4}
\usepackage{graphicx}
\usepackage[dvipdfm,bookmarks=true]{hyperref}

\begin{document}

\title{\LARGE{\bf  Dielectron and Diphoton channels in 2HDM}}
\date{\today}
\author{Mingxing Luo}
\email{luo@zimp.zju.edu.cn}
\author{Liucheng Wang}
\email{liuchengwang@zimp.zju.edu.cn}
\author{Guohuai Zhu}
\email{zhugh@zju.edu.cn}
\affiliation{Zhejiang Institute of Modern
Physics, Department of Physics, Zhejiang University, Hangzhou,
Zhejiang 310027, P.R.China}

\begin{abstract}
It was observed recently that \cite{Randall},
any particle in the Standard Model cannot decay to $e^+e^-$ and $\gamma \gamma$
final states with comparably measurable branching ratios.
This is also true for most extensions of the Standard Model, with the Randall-Sundrum model as an outstanding exception.
In this paper, we show that two-Higgs-Doublet-Models (2HDM) yield another possible exception
if certain parameters are properly chosen. In addition,
we have checked that this model survives the tests of low energy processes, including the anomalous magnetic moment and
electric dipole moment of leptons, lepton-flavor-violating decays $\mu^- \to e^- \gamma$ and $e^- e^+ e^-$.
\end{abstract}

\maketitle

\section{Introduction}
It was observed recently that \cite{Randall},
no resonance can decay to $e^+e^-$ and $\gamma \gamma$ with comparably measurable branching ratios,
if electrons interact only with the electroweak gauge bosons and the Standard Model (SM) Higgs boson.
That is to say,
if both dielectron and diphoton channels are observed in experiments such as Tevatron and/or LHC,
one gets a smoking-gun evidence of new physics beyond the SM.
Furthermore, this observation holds true for most conventional extensions of the SM,
with the Randall-Sundrum (RS) \cite{RS} model as an exception.
Specifically, the ratio $Br(h^{KK} \to \gamma \gamma)/Br(h^{KK} \to e^+ e^-)$
for the Kaluza-Klein (KK) modes $h^{KK}$ of graviton is predicted to be exactly $2$ in the original RS-model.
In variant RS-models, this ratio could be different,
but generically one expects to observe simultaneously $e^+e^-$ and $\gamma \gamma$ final states with comparable rates.
We are thus led to examine carefully whether there are other extensions of the SM
which can produce similar signals.

First, let us recapitulate the arguments \cite{Randall}
why $e^+ e^-$ and $\gamma \gamma$ channels do not appear simultaneously in the SM.
Due to conservation of angular momentum,
fermionic resonance can decay to neither $e^+e^-$ nor $\gamma\gamma$ final states.
Furthermore, Landau-Yang theorem \cite{Landau-Yang} prohibits a spin-1 resonance decaying into two photons,
which follows from general assumptions of Lorentz invariance, gauge invariance, and Bose symmetry of the photons.
In principle, a spin-0 resonance can decay to $e^+ e^-$
either through the electroweak gauge interaction or the Yukawa interaction.
However, the gauge channel is helicity suppressed
while the Yukawa channel is suppressed by the ratio of electron mass to the vacuum expectation value (VEV) of the SM Higgs.
For a resonance of spin-2 or higher spin,
$e^+ e^-$ channel is forbidden at tree level simply
because the resonance can not couple directly to a virtual spin-1 gauge boson.
The arguments hold also true for the case of dimuon and diphoton signals,
if the muon mass is much smaller than the energy scale of the resonance.

Thus comparably measurable dielectron and diphoton signals clearly imply that leptons must have direct interaction
with either a spin-2 (or higher spin) boson or a new scalar beyond the SM Higgs.
The former possibility has been examined in \cite{Randall} where one gets the KK modes of RS graviton as a concrete example.
This is actually not hard to understand,
as gravity couples universally to energy-momentum tensor of any fields, including electrons and photons.
In this letter, we will instead focus on the latter case, by considering models with new scalar fields
that can decay into dielectron and diphoton with comparably measurable rates.
In particular, we will concentrate on extensions of the SM with an extra Higgs doublet,
which is usually dubbed as the two Higgs doublet models (2HDM).

In 2HDM, there are eight degrees of freedom in the scalar sector to start with.
After spontaneously symmetry breaking,
five scalars will remain but two of them are charged.
Among the three neutral ones, one resembles the SM Higgs
and the additional two could play the role of new resonances.
These neutral scalars couple to electron directly through Yukawa interactions.
To avoid constraints on the Yukawa couplings due to the light electron mass compared to the electroweak scale,
the VEV of the additional Higgs doublet must be (almost) zero.
This type of model is usually referred to as type III 2HDM,
in which the neutral scalars can decay to $e^+ e^-$ at tree level with significant rate.
Of course, care should be taken to avoid large FCNC and CP violation effects.
On the other hand, the $\gamma \gamma$ decay is loop suppressed in these models
and smaller than that in the SM due to the absence of contributions from gauge particles.
To make a sizable diphoton decay rate, some mechanism for enhancement is needed for this loop level process.
In the letter, we will see that such enhancement is still feasible within current experimental limits.
Putting all things together,
the extra scalars can be made to decay into both $e^+e^-$ and $\gamma \gamma$
final states with comparably measurable branching ratios.
As one would have expected, here one has to choose parameters properly and certain level of fine tuning is necessary.

This paper is organized as follows: In Section II, we will show how to
get significantly comparable branching ratios for $e^+ e^-$ and $\gamma \gamma$ channels in 2HDM.
Phenomenological constraints on these models are presented in Section III,
including the anomalous magnetic moment of leptons, electric dipole
moment(EDM) of leptons, the lepton flavor violating (LFV) processes $\mu^- \rightarrow e^- \gamma$ and
$\mu^- \rightarrow e^- e^+ e^-$.
Generically leptonic Yukawa couplings in 2HDM are mostly constrained by the electric dipole moment (EDM) of the electron,
instead of by the LFV processes.
The results are summarized in section IV.

\section{Comparable branching ratios in Type III 2HDM}

In Type III 2HDM, one scalar doublet resembles the role of the SM Higgs boson, which will be denoted as $\phi_1$.
We denoted the additional doublet as $\phi_2$.
The general interaction between $\phi_2$ and fermions reads
\begin{eqnarray}
{\cal{L}}_{Yukawa}&=&- \xi^U_{ij}\bar{Q}_{Li} \tilde{\phi_{2}} U_{R
j}-\xi^D_{ij}\bar{Q}_{Li} \phi_{2}D_{Rj}-\xi^E_{ij}\bar{l}_{Li}
\phi_{2}E_{Rj}+ h.c. \label{lagrangian}
\end{eqnarray}
with left (right) handed projection $L(R)=(1\mp
\gamma_5)/2$, $\bar{Q}_{Li}$ ($l_{Li}$) are left handed quark (lepton) doublets, $U_{Rj}$,
$D_{Rj}$ and $E_{Rj}$ are right handed up quark, down quark and lepton singlets, respectively, with family indices $i,j$.
In general, all Yukawa matrices $\xi^{U,D,E}_{ij}$ are non-diagonal and complex\footnote
{The Yukawa matrices for interactions between $\phi_1$ and fermions are diagonalized here,
so all fermions are already in their mass eigenstates.}.
This may result in FCNC and CP violating effects, which will be addressed in the next section.
The general Higgs potential \cite{Guide}
which spontaneously breaks $SU(2)_L\times U(1)_Y$ down to $U(1)_{EM}$ in 2HDM is
\begin{eqnarray}
V(\phi_1, \phi_2 )&=&  \lambda_1 (\phi_1^+
\phi_1-v^2_1)^2+ \lambda_2 (\phi_2^+ \phi_2-v^2_2)^2 + \lambda_3
[(\phi_1^+ \phi_1-v^2_1)+ (\phi_2^+ \phi_2-v^2_2)]^2 \nonumber \\
&+& \lambda_4 [(\phi_1^+ \phi_1) (\phi_2^+ \phi_2)-(\phi_1^+
\phi_2)(\phi_2^+ \phi_1)]+ \lambda_5 [Re(\phi_1^+
\phi_2)-v_1v_2\cos\xi]^2 \nonumber
\\ &+& \lambda_{6} [Im(\phi_1^+ \phi_2)-v_1v_2\sin\xi]^2
 , \label{potential}
\end{eqnarray}
where the $\lambda_i$'s are all real parameters.

In type I and type II 2HDM, the non-zero VEVs of $\phi_1$ and $\phi_2$ make contributions to fermion masses
proportional to the corresponding Yukawa couplings in (\ref{lagrangian}).
The branching ratio of the $e^+e^-$ final state from scalar particles
is severely suppressed by the tiny electron mass.
In type III 2HDM, those Yukawa couplings are not confined by such constraints,
since $\phi_2$ has a (almost) zero VEV.
One might hope to find viable models with comparably measurable branching ratios
for $e^+ e^-$ and $\gamma \gamma$ channels.

Only $\phi_{1}$ has a non-zero VEV while $\phi_{2}$ does not in type III 2HDM.
That is,
\begin{eqnarray}
v_1=<\phi_{1}>=\frac{1}{\sqrt{2}}\left(\begin{array}{c c}
0\\v\end{array}\right) \,  \, ; \hspace*{1.7cm} v_2= <\phi_{2}>=0 \,\, . \label{vev}
\end{eqnarray}
We can thus choose a gauge such that
\begin{eqnarray}
\phi_{1}=\frac{1}{\sqrt{2}}\left(\begin{array}{c c}
0\\v+h\end{array}\right) \,  \,; \hspace*{1.7cm}
\phi_{2}=\frac{1}{\sqrt{2}}\left(\begin{array}{c c} \sqrt{2} H^{+}\\
H_1+i H_2 \end{array}\right) \,\, , \label{define}
\end{eqnarray}
In this gauge, the potential in (\ref{potential}) can be re-expressed as,
\begin{eqnarray}
V(\phi_1, \phi_2
)&=&\frac{1}{2}\left [2(\lambda_1+\lambda_3)v^2 \right ]h^2+\left (\frac{1}{2}\lambda_4
v^2 \right )H^-H^++\frac{1}{2}\left (\frac{1}{2}\lambda_5 v^2 \right )H^2_1\nonumber
\\&+&\frac{1}{2}\left (\frac{1}{2}\lambda_6 v^2\right )H^2_2+\textrm{(interaction terms)}\, ,\label{potential2}
\end{eqnarray}
One sees that $H_1,H_2$ and $h$ are mass eigenstates, i.e., there is no mixing between the Higgs bosons.
To be definite, $H_1$ is always assumed to be lighter than $H_2$, without losing any generality.
Now we will see how to make $H_1$ decaying into both $e^+ e^-$ and $\gamma \gamma$ channels with comparable
and significant rates.

\begin{figure}
\centerline{\includegraphics[height=5cm]{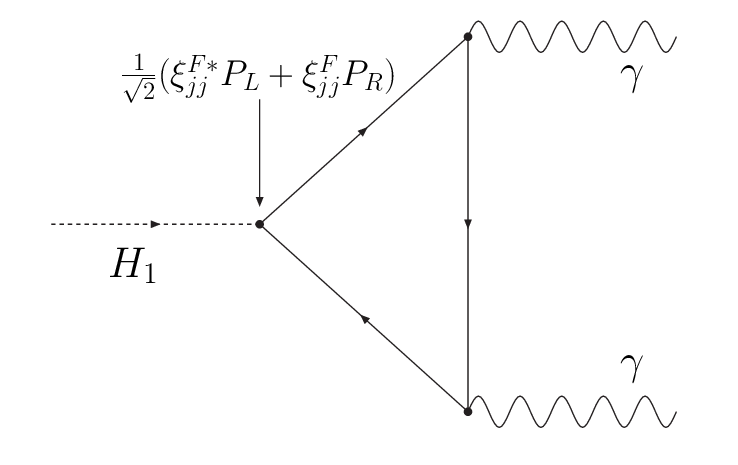}}
\caption{$H_1 \to \gamma\gamma$ at one-loop level. The internal lines represent either leptons or quarks.
}\label{fig_higgs_to_two_gamma}
\end{figure}
It is clear from (\ref{lagrangian}) that $H_1$ decays into $e^+e^-$ at the tree level.
By ignoring the electron mass, the decay width is
\begin{eqnarray}
\Gamma(H_1\rightarrow
e^{+} e^{-})=\frac{M_{H_1}}{16\pi}\xi^E_{ee}\xi^{E*}_{ee}\,
,\label{h1to2gamma}
\end{eqnarray}
where $M_{H_1}$ is the mass of $H_1$ and $\xi^E_{ee}$ is the Yukawa coupling between $H_1$ and electron.
Meanwhile, $H_1$ decays to $\gamma \gamma$ through triangle fermion loop (see Fig. (\ref{fig_higgs_to_two_gamma})).
Notice that the triangle gauge boson loops, which are dominant in the SM, are absent here.
This is because that the term $(D_\mu \phi_2)^\dag (D_\mu \phi_2)$ only results in a
four-point vertex $\phi^+_2 \phi_2 B^\mu B_\mu$ due to the vanishing vev of $\phi_2$.
Similarly the scalar loops do not appear here either.
Fig. (\ref{fig_higgs_to_two_gamma}) yields the following decay amplitude,
\begin{eqnarray}
M(H_1\rightarrow\gamma \gamma)&=&
2A^+(pqg^{\mu\nu}-p^{\nu}q^{\mu})\varepsilon^*_{\mu}(p)\varepsilon^*_{\nu}(q)-
2A^-i\epsilon^{\mu\nu\rho\sigma}p_{\rho}q_{\sigma}\varepsilon^*_{\mu}(p)\varepsilon^*_{\nu}(q)\nonumber
\\&=&
A^+F_{\mu\nu}F^{\mu\nu}+iA^-F_{\mu\nu}\widetilde{F^{\mu\nu}}\,,
\end{eqnarray}
with
$\widetilde{F^{\mu\nu}}=\frac{1}{2}\epsilon^{\mu\nu\rho\sigma}F_{\rho\sigma}$.
There are both CP even amplitude $A^+$ and CP odd amplitude $A^-$,
as the Yukawa couplings are complex,
\begin{eqnarray}
A^+ &=& \sum\limits_{j}\frac{-ie^2Q^2_j
M_j}{8\pi^2M^2_{H_1}}(\frac{\xi^F_{jj}}{\sqrt2}+\frac{\xi^{F*}_{jj}}{\sqrt2})\int_{0}^{1}dx\int_{0}^{1-x}dy\frac{1-4xy}{t_j-xy}
\nonumber
\\A^- &=& \sum\limits_{j}\frac{-ie^2Q^2_j M_j}{8\pi^2M^2_{H_1}}(\frac{\xi^F_{jj}}{\sqrt2}-\frac{\xi^{F*}_{jj}}{\sqrt2})\int_{0}^{1}dx\int_{0}^{1-x}dy\frac{1}{t_j-xy}
\end{eqnarray}
where $t_j={M^2_j}/{M^2_{H_1}}$, $F=E,U$ or $D$ for the virtual fermions to be charged leptons,
up quarks or down quarks. $Q_j$ are their charges.
The decay width of $H_1\rightarrow \gamma \gamma$ is thus,
\begin{eqnarray}
\Gamma(H_1\rightarrow \gamma \gamma)=\frac{M^3_{H_1}}{16
\pi}(|A^+|^2+|A^-|^2)\label{h1to2e}
\end{eqnarray}
which agree with results in \cite{2gamma}.

Now we are in a position to make $H_1$ decays into $e^+ e^-$ and $\gamma \gamma$ with comparable and significant rate.
To enhance the loop-suppressed $\gamma \gamma$ channel,
the Yukawa couplings of $H_1$ to heavy leptons and/or quarks should be much larger than that of $H_1$ to electron.
Let us first consider case of a quark-phobic $\phi_2$.
We need then a heavy lepton to have a large coupling with $\phi_2$,
which we may, for example, choose it to be the $\tau$ lepton.
Taking the mass of $H_1$ to be a few hundred GeV,  $|\xi^E_{\tau\tau}|/|\xi^E_{ee}| > 10^3$ is required
to make the branching ratios for $e^+ e^-$ and $\gamma \gamma$ comparable.
However with this parameter set, the $e^+ e^-$ channel would be too rare to
be observable:
\begin{eqnarray}
{\rm
Br}(H_1\rightarrow e^+e^-)<\frac{\Gamma(H_1\rightarrow
e^+e^-)}{\Gamma(H_1\rightarrow
\tau^+\tau^-)}\sim\frac{|\xi^E_{ee}|^2}{|\xi^E_{\tau\tau}|^2} < 10^{-6}
\end{eqnarray}
So a quark-phobic $\phi_2$ will not work.

At first glance, the same would happen if $\phi_2$ has large couplings to quarks.
Naively, ${\rm Br}(H_1\rightarrow e^+e^-) \ll {\rm Br}(H_1\rightarrow q \bar{q})$ would make $e^+ e^-$ final state
too scarce to be observed.
Fortunately, if the mass of $H_1$ is smaller than $340$ GeV, roughly twice of top quark mass,
$H_1$ cannot decay to top quark pairs.
In this case, one can make the $e^+ e^-$ and $\gamma \gamma$ decay channels comparable and observable,
by having $\xi^U_{tt} \gg \xi^E_{ee}$ and keeping the Yukawa couplings between $\phi_2$ and the other quarks small.
$H_1$ cannot be too light either, otherwise it should have been observed at LEP already,
as the coupling between $H_1$ and electrons is not too small in this model.
For illustrations, we will take (somewhat {\it ad hoc}) $M_{H_1}=200~\mbox{GeV}$
and $M_{H_2}=M_{H^+}=300~\mbox{GeV}$ in the following numerical discussions.

To get $\Gamma(H_1\rightarrow e^+e^-)$/$\Gamma(H_1\rightarrow \gamma \gamma) \simeq 1$,
one needs the ratio of Yukawa couplings to be $\xi^U_{tt}/\xi^E_{ee} \simeq 1800$.
To make $e^+ e^-$ and $\gamma \gamma$ decay channels observable,
we may assume universal coupling of $H_1$ to all fermions except the top quark\footnote
{The Yukawa couplings between $\phi_2$ and other fermions need only to be of the same order of $\xi^E_{ee}$.
A universal coupling is not necessary, though it makes numerical analysis simpler.}.
With this parameter set, the main decay channel will be $H_1 \rightarrow gg$.
It is then easy to estimate the branching ratios:
\begin{eqnarray}
 {\rm Br}(H_1\rightarrow
e^+ e^-)\simeq {\rm Br}(H_1\rightarrow
\gamma \gamma)\sim\frac{\Gamma(H_1\rightarrow
\gamma \gamma)}{\Gamma(H_1\rightarrow
g g)}\sim\frac{\alpha^2}{2\alpha^2_S}\simeq 1/200~,
\end{eqnarray}
which is not too small and should be measurable.
Therefore with such a specific parameter set,
it would be possible for Type III 2HDM to produce significant and comparable
branching ratios for $e^+e^-$ and $\gamma\gamma$ decays.

In this parameter set, the production of $H_1$ at hadron collider is
mainly through gluon fusion, which is similar to that of the SM Higgs.
Therefore, to copiously produce $H_1$ at an observable rate, one would expect
$\xi^U_{tt}$ to be of order one.
Generically, one needs $\xi^E_{ee}$ to be no less than $10^{-4}$.
In the following, we will discuss whether this is possible
concerning phenomenological constraints from low energy processes.

\section{Phenomenological constraints and Implications}

With extra not-so-heavy Higgs particles, the type III 2HDM has rich phenomenological implications at low energy.
For example, it may contribute to the anomalous magnetic moments of leptons,
which have been measured to very high precision.
Complex phases of the Yukawa couplings of $H_1$ to leptons may contribute
to leptonic EDMs at one-loop level.
Finally, non-diagonal Yukawa couplings of $H_1$ to leptons would lead to LFV processes,
such as $\mu \to e\gamma$ and $\mu^- \to e^- e^+ e^-$.
Non-diagonal Yukawa couplings of $H_1$ to quarks might produce large hadronic FCNC processes
and complex phases may induce CP violating effects in hadrons,
which are both tightly constrained experimentally.
However these couplings do not affect $e^+ e^-$ and $\gamma \gamma$ signals at all,
therefore we will simply assume that their effects can be tuned away and not consider them further.

The complex Yukawa matrix for leptonic couplings $\xi^E_{ij}$ contains $18$ real parameters,
which are obviously too many.
So we will consider two extreme scenarios, just for illustration, in numerical discussions:

Scenario I (S1): the leptonic Yukawa matrix is diagonal with a universal CP phase $\pi/6$,
i.e. $\xi^E_{ij}=|\xi_{ij}|e^{i\theta_{ij}}$, with $|\xi_{ij}|=0$(for
$i\neq j$) and $\theta_{ii}=\pi/6$(for $i=e,\mu,\tau$).

Scenario II (S2) all leptonic Yukawa couplings are assumed to have universal magnitude with CP phase $\pi/6$,
namely $\xi^E_{ij}=\xi e^{i \pi/6}$.

Actually, numerical analyses in these two scenarios yield rather reasonable estimates
on constraints on the general Yukawa matrix.

\subsection{$g-2$}
\begin{figure}
\centerline{\includegraphics[height=5cm]{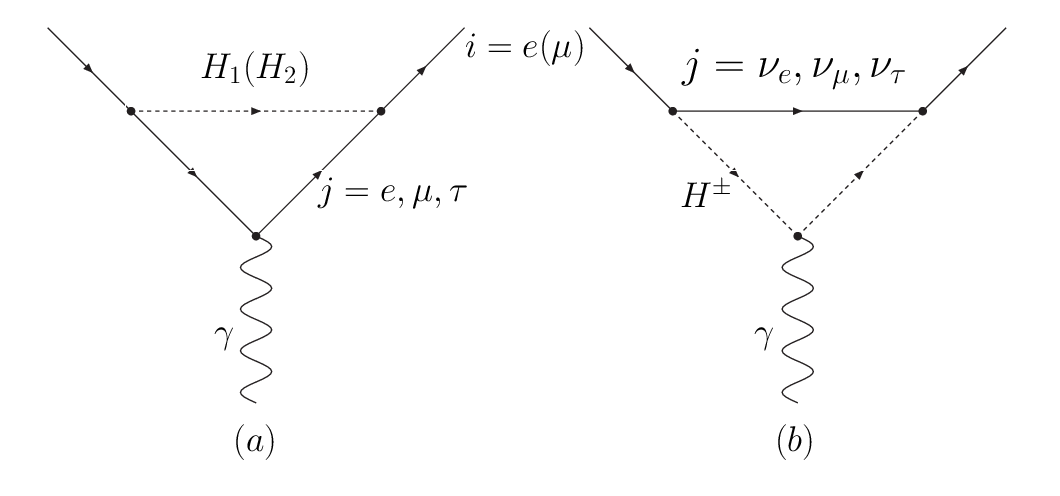}} \caption{The one-loop
diagrams contribute to g-2 and EDM in type III 2HDM.
}\label{fig_g-2_EDM}
\end{figure}

For the electron, the experimental data gives $a_e^{exp} \equiv (g_e-2)/2 = (1159652181.1
\pm 0.7) \times 10^{-12}$ \cite{Data} which agrees well with the SM calculation
$a_e^{SM}=(1159652182.8 \pm 7.7) \times 10^{-12}$\cite{g-2}. The extra scalars in 2HDM may
contribute to the anomalous magnetic moment at one-loop level, as shown in Fig. (\ref{fig_g-2_EDM}).
We will assume $a_e^{2HDM} < 10 \times 10^{-12}$ which is slightly larger than the theoretical error,
to get a restriction on the corresponding Yukawa couplings.
For muon, there is some discrepancy
between experiments and the SM calculation \cite{Data}:
$a_\mu^{exp}-a_\mu^{SM} = (292 \pm 63 \pm 58) \times 10^{-11}$, so we assume an upper limit
$a_\mu^{2HDM} < 292 \times 10^{-11}$.

Fig. (\ref{fig_g-2_EDM}) yields the one-loop contribution to $g-2$ via neutral and charged Higgs as
\begin{eqnarray}
a_i &=&
\sum\limits_{j=e,\mu,\tau}
\left[ \frac{1}{16\pi^2}\int_{0}^{1}dz\frac{-\xi^E_{ji}\xi^{E*}_{ji}M^2_iz(1-z)}{M^2_{H^+}-zM^2_i-i\epsilon}
\right. \nonumber
\\&+&\frac{1}{32\pi^2}\int_{0}^{1}dz\frac{M^2_iz^2(1-z)(\xi^E_{ij}\xi^{E*}_{ij}
+\xi^{E*}_{ji}\xi^E_{ji})+M_iM_jz^2(\xi^{E*}_{ji}\xi^{E*}_{ij}+\xi^E_{ij}\xi^E_{ji})}{(1-z)M^2_{H_1}
+zM^2_j-z(1-z)M^2_i-i\epsilon}\nonumber
\\&+&
\left. \frac{1}{32\pi^2}\int_{0}^{1}dz\frac{M^2_iz^2(1-z)(\xi^E_{ij}\xi^{E*}_{ij}
+\xi^{E*}_{ji}\xi^E_{ji})-M_iM_jz^2(\xi^{E*}_{ji}\xi^{E*}_{ij}+\xi^E_{ij}\xi^E_{ji})}{(1-z)M^2_{H_2}
+zM^2_j-z(1-z)M^2_i-i\epsilon} \right]\label{eq_g-2}
\end{eqnarray}
with $i=e,\mu$.

In S1, the expression (\ref{eq_g-2}) is proportional to $M_e^2$ ($M_\mu^2$) for
electron (muon) anomalous magnetic moment. That is why, though the upper limit on $a_{\mu}^{2HDM}$
is about $300$ times larger than that of $a_e^{2HDM}$, the constraint $\xi_{\mu \mu} <0.64$ from
$a_\mu^{2HDM} < 292 \times 10^{-11}$ is more severe than the limit $\xi_{ee}< 5.9$ from $a_e^{2HDM} < 10
\times 10^{-12}$.

In S2, the expression (\ref{eq_g-2}) is dominated by $M_eM_\tau$ ($M_\mu M_\tau$) term for
the electron (muon) anomalous magnetic moment.
In this scenario, the electron $g-2$ leads to $\xi< 0.18$, which is
slightly smaller than the restriction $\xi< 0.21$ from the muon $g-2$.

\subsection{EDM}

The electron and muon EDMs are extremely small in the SM, which arise at three-loop level.
But in the present model, the complex Yukawa couplings give rise to non-zero EDMs at one loop level,
as shown in Fig.(\ref{fig_g-2_EDM}). Thus these couplings might be severely constrained by
the experimental data.
The current experimental measurements cited by PDG2008 \cite{Data} are
 $d_e^{exp}= (0.07 \pm 0.07) \times 10^{-26}$ e cm and
$d_\mu^{exp}= (3.7 \pm 3.4) \times 10^{-19}$ e cm. The corresponding upper limits can be found in the original
experimental literature: $d_e < 0.16 \times 10^{-26}$ e cm
(90\% C.L.)\cite{EDM1} and $d_\mu < 10 \times 10^{-19}$ e cm
(95\% C.L.)\cite{EDM2}. We will use these upper limits, instead of the PDG numbers, to
obtain constraints. The leptonic EDM $d_i$
($i=e,\mu$) has contributions from neutral Higgs bosons $H_1$, $H_2$ and charged one $H^+$.
The contribution from the internal $H^+$ is suppressed by the neutrino masses, which can be safely dropped.
The contribution from the neutral Higgs bosons is
\begin{eqnarray}
d_i &=&
\sum\limits_{j=e,\mu,\tau}\frac{ie}{64\pi^2}\int_{0}^{1}dz
\left[ \frac{M_jz^2(\xi^E_{ij}\xi^E_{ji}-\xi^{E*}_{ji}\xi^{E*}_{ij})}{(1-z)M^2_{H_1}+zM^2_j-z(1-z)M^2_i-i\epsilon}
\right. \nonumber \\&&
~~~~~~~~~~~~~~~~~~~~~~~~~
\left. -\frac{M_jz^2(\xi^E_{ij}\xi^E_{ji}-\xi^{E*}_{ji}\xi^{E*}_{ij})}{(1-z)M^2_{H_2}+zM^2_j-z(1-z)M^2_i-i\epsilon}
\right] \,,\label{eq_EDM}
\end{eqnarray}
As expected, $d_i$ vanishes when the phase of $\xi^E_{ij}$ is zero and
will not yield any constraints on the Yukawa couplings.
If only virtual $\tau$ lepton is considered inside the loop and
the external lepton mass is neglected, our formula agrees with \cite{EDM}. The above equation
is also consistent with \cite{Review}.

In S1, the expression
(\ref{eq_EDM}) is proportional to $M_e$ ($M_\mu$) for electron
(muon) EDM. It is then straightforward to find $\xi_{ee}< 0.013$ and $\xi_{\mu \mu} <31$
from the experimental upper limits.

In S2, the expression (\ref{eq_EDM}) is dominated by
$M_\tau$ term for both $d_e$ and $d_\mu$. Therefore the constraint $\xi< 3.9 \times 10^{-4}$ from
$d_e$ is significantly tighter than the limit $\xi< 9.6$ from $d_\mu$.

\subsection{LFV processes: $\mu^- \rightarrow e^- \gamma$ and $e^- e^+ e^-$}

\begin{figure}
\centerline{\includegraphics[height=5cm]{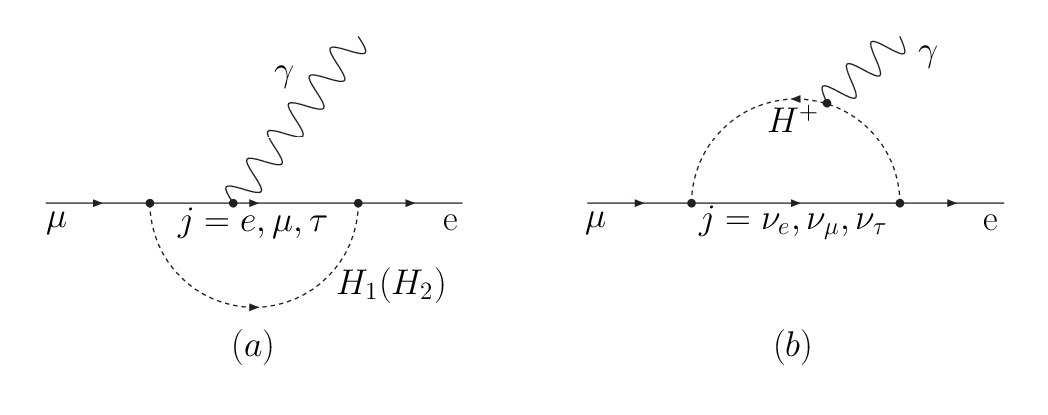}}
\caption{$\mu\rightarrow e\gamma$ decay in 2HDM at one-loop level.}
\label{fig_mu_to_e_and_gamma}
\end{figure}

Both neutral scalars $H_{1,2}$ and charged one $H^+$ can contribute to $\mu \to e \gamma$ decay
through loop diagrams, as shown in Fig. (\ref{fig_mu_to_e_and_gamma}).
This channel has not been observed, but with an upper limit ${\rm Br}(\mu^-
\to e^- \gamma)<1.2 \times 10^{-11}$ \cite{Data}. For muon decays into $e^- e^+ e^-$ final state,
it may even occur at tree level in our model. Experimentally, the upper limit
${\rm Br}(\mu^- \to e^- e^+ e^- )<1.0 \times 10^{-12}$ \cite{Data} is also very small.

The decay amplitude of $\mu \to e \gamma$ can be expressed as
\begin{eqnarray}
M(\mu \rightarrow
e+\gamma)=\frac{e\varepsilon^*_{\mu}(q)}{16\pi^2}(-i)\overline{U}_e(p\prime)(p+p\prime)^{\mu}(A_LP_L+A_RP_R)U_{\mu}(p)~,
\end{eqnarray}
where $A_L$ and $A_R$ are,
\begin{eqnarray}
A_L &=&
\sum\limits_{j=e,\mu,\tau}\int_{0}^{1}dx\int_{0}^{1}dy\int_{0}^{1}dz\delta(x+y+z-1)\nonumber
\\&&\left[ \frac{xzM_{\mu}\xi^{E*}_{je}\xi^E_{j\mu}}{(1-z)M^2_{H^+}-xzM^2_{\mu}-yzM^2_e}+\frac{-\frac{1}{2}yzM_{\mu}\xi^E_{ej}\xi^{E*}_{\mu
j}
-\frac{1}{2}xzM_e\xi^{E*}_{je}\xi^E_{j\mu}-\frac{1}{2}M_j(1-z)\xi^{E*}_{je}\xi^{E*}_{\mu
j}}{zM^2_{H_1}+(1-z)M^2_j-yzM^2_\mu-xzM^2_e} \right. \nonumber\\
&& \left.~~ + \frac{-\frac{1}{2}yzM_{\mu}\xi^E_{ej}\xi^{E*}_{\mu
j}-\frac{1}{2}xzM_e\xi^{E*}_{je}\xi^E_{j\mu}+\frac{1}{2}M_j(1-z)\xi^{E*}_{je}\xi^{E*}_{\mu
j}}{zM^2_{H_2}+(1-z)M^2_j-yzM^2_{\mu}-xzM^2_e} \right]~,\nonumber
\\A_R &=& \sum\limits_{j=e,\mu,\tau}\int_{0}^{1}dx\int_{0}^{1}dy\int_{0}^{1}dz\delta(x+y+z-1)\nonumber
\\&& \left[ \frac{yzM_e\xi^{E*}_{je}\xi^E_{j\mu}}{(1-z)M^2_{H^+}-xzM^2_{\mu}-yzM^2_e}+\frac{-\frac{1}{2}yzM_{\mu}\xi^E_{j
\mu}\xi^{E*}_{je}-\frac{1}{2}xzM_e\xi^{E*}_{\mu
j}\xi^E_{ej}-\frac{1}{2}M_j(1-z)\xi^E_{ej}\xi^E_{j\mu}}{zM^2_{H_1}+(1-z)M^2_j-yzM^2_{\mu}-xzM^2_e} \right. \nonumber
\\&&\left. ~~ +\frac{-\frac{1}{2}yzM_{\mu}\xi^E_{j\mu}\xi^{E*}_{je}-\frac{1}{2}xzM_e\xi^{E*}_{\mu j}\xi^E_{ej}+\frac{1}{2}M_j(1-z)\xi^E_{ej}\xi^E_{j\mu}}{zM^2_{H_2}+(1-z)M^2_j-yzM^2_{\mu}-xzM^2_e} \right]~.
\end{eqnarray}
In terms of $A_{L,R}$, the decay width can be expressed as,
\begin{eqnarray}
\Gamma(\mu \rightarrow e+\gamma)=\frac{\alpha M^3_{\mu}}{1024
\pi^4}(|A_L|^2+|A_R|^2)~.
\end{eqnarray}

For $e^- e^+ e^-$ channel, ignoring the mass of electron, the decay width can be calculated as
\begin{eqnarray}
\Gamma=\frac{|\xi^E_{ee}|^2\xi^{E*}_{\mu e}\xi^E_{e \mu}M^5_{\mu}}{24576
\pi^3 }\left ( \frac{1}{M^4_{H_1}}+ \frac{1}{M^4_{H_2}}\right )\,.
\end{eqnarray}

In S1, these LFV processes are forbidden because of the diagonal lepton Yukawa coupling.
In S2, we find numerically the restriction $\xi<1.2\times 10^{-3}$ from the $e \gamma$ channel,
which is smaller than the constraint $\xi<2.2\times 10^{-3}$ from the $e^- e^+ e^-$ channel.

\section{Summary}\label{conclusions}
We have so far found that,
type III 2HDM may produce comparably observable $e^+e^-$ and $\gamma \gamma$ signals simultaneously from
decays of neutral scalars.
This provides an interesting alternative to the RS model.
To make this happen, certain amount of fine tunings are necessary.
First, to enhance the loop suppressed $\gamma \gamma$ channel to be comparable with $e^+ e^-$ channel,
the Yukawa coupling of $H_1$ to top quark ($\xi^U_{tt}$) must be a few thousand times larger than
those to the rest of fermions.
In addition, the mass of $H_1$ should be smaller than twice of the top quark mass,
to prevent $H_1$ decays predominantly to top-anti-top pair.
Under these conditions, $H_1$ would decay mainly to two gluons,
with the branching ratios of $e^+ e^-$ and $\gamma \gamma$ channels to be about half a percent.

Notice that these requirements determine only the relative strength of different Yukawa couplings,
while their absolute sizes would determine the number of events produced in colliders.
If the parameters in these models do meet the above requirements,
$H_1$ will be mainly produced via gluon fusion in hadron colliders, analogous to the SM Higgs.
For $H_1$ to be produced roughly at the same rate as the SM Higgs,
$\xi^U_{tt}$ needs to be of order one,
which means that the electron Yukawa coupling $\xi^E_{ee}$ should be around few $\times 10^{-4}$.

Generically, the non-diagonal and complex Yukawa couplings lead to FCNC and CP violating processes.
We have thus examined constraints from the anomalous magnetic moment and EDM of leptons,
LFV processes $\mu^- \to e^- \gamma$ and $e^- e^+ e^-$.
Because of the large number (18 in total) of free parameters in Yukawa interactions,
we have chosen two scenarios as illustration for numerical discussions.
In the scenario of diagonal Yukawa couplings with an universal CP phase, we find the most stringent constraint
$\xi^E_{ee}<0.013$ from electron EDM, and $\xi^E_{\mu \mu}<0.64$ from muon $g-2$ measurement.
In the scenario of universal Yukawa couplings, the most stringent constraint $\xi < 3.9 \times 10^{-4}$
comes again from the electron EDM.
In both scenarios, the constraint on $\xi^E_{ee}$ is above the requirement from the dielectron and diphoton signals,
so the model survives existing phenomenological tests.
The same conclusion also holds true for general Yukawa interactions.

So, if resonances below $2M_t$ are found in experiments to decay into $e^+e^-$ and $\gamma \gamma$
with comparable rates, more work would be needed to delineate their origins.
Under this circumstance, we have just seen that type III 2HDM provides one alternative possibility,
in addition to the RS-model.
It would be interesting to see whether there exist other possibilities.

\begin{acknowledgments}
We would like to thank Mark Wise for valuable comments on the manuscript.
This work is supported in part by the National Science Foundation of
China under grant No.10425525, No.10875103 and No.10705024. G.Z is also supported
in part by the Scientific Research Foundation for the Returned Overseas
Chinese Scholars, State Education Ministry.
\end{acknowledgments}

\end{document}